\documentclass[journal]{IEEEtran}
\usepackage[cmex10]{amsmath}
\usepackage{array}

\ifCLASSINFOpdf
   \usepackage[pdftex]{graphicx}
\else
\fi

\begin{document}

\title{Experimental near field OAM-based communication with circular patch array}
\author{Fabio~Spinello,~Elettra~Mari,~Matteo~Oldoni,~Roberto~A.~Ravanelli,~Carlo~G.~Someda\\~Fabrizio~Tamburini,~Filippo~Romanato,~Piero~Coassini~and~Giuseppe~Parisi
\thanks{F. Spinello is with Department of Information Engineering, University of Padova, via Gradenigo 6B, I-35131 Padova, Italy, e-mail: spinello@dei.unipd.it.}
\thanks{E.~Mari,~G. Parisi, C. G. Someda, F. Tamburini and F. Romanato are with Twist Off s.r.l., via della Croce Rossa 112, I-35129 Padova, Italy.}
\thanks{M. Oldoni, R. A. Ravanelli and P. Coassini are with SIAE Microelettronica, via Michelangelo Buonarroti 21, I-20093 Cologno Monzese, Milan, Italy.}
\thanks{F. Spinello is also with Twist Off s.r.l., via della Croce Rossa 112, I-35129 Padova, Italy.}
\thanks{F. Romanato is also with Department of Physics and Astronomy, University of Padova, via Marzolo 8, I-35100 Padova, Italy.}}

\maketitle

\begin{abstract}
A short range experimental communication system, based on Orbital Angular Momentum (OAM) multiplexing, is presented. We characterize circular arrays of patch antennas designed to transmit and receive OAM electromagnetic fields, reporting new results on communication links based on such antennas.
We also experimentally study the antennas tolerance to  misalignment errors (angular tilt and lateral shift) within which OAM multiplexing can be efficiently exploited. Starting from these results, we finally propose an application to short range communications of OAM-based systems that can lead to a high level of security in the information exchange.

\end{abstract}

\begin{IEEEkeywords}
orbital angular momentum, communication, radio
\end{IEEEkeywords}

\IEEEpeerreviewmaketitle

\section{Introduction}
Electromagnetic (EM) fields can carry not only energy, but also spin angular momentum and orbital angular momentum (OAM) \cite{allen1992orbital}. 
The properties of OAM have attracted an increasing interest in different research fields in the last two decades \cite{yao2011orbital}, bringing to several applications in applied physics \cite{furhapter2005spiral, mari2012sub, wang2012terabit}. Recently, the study of OAM has been extended also to radio domain \cite{thide2007utilization}, in particular, to radio telecommunication for the possibility of implementing channel multiplexing and frequency reuse, as demonstrated in optics \cite{huang2014100}.
In fact, OAM waves do not interact during propagation in a homogeneous medium, i.e., they form an orthogonal set of propagating modes \cite{mari2015near}. 
The exploitation of OAM in communication links \cite{tamburini2012encoding, tamburini2015tripling} has opened a discussion about the actual possibility of overcoming some constraints, like the effect of OAM modes propagation or the  misalignment between transmitting and receiving antennas \cite{edfors2012orbital}. This debate is restricted to long-distance transmission.
Whereas, no particular restrictions on the use of OAM waves on short distances are evident. 
So, in near-field range, OAM modes could represent a way to implement channel multiplexing and their natural orthogonality would simplify signals processing. 
Also, OAM modes could implement a multi-channel high-rate data link with low complexity processing and wireless applications for communicating within a data center or within a server farm. The study of such applications is fairly recent. Some authors started with theoretical analysis, looking in particular at the systems performances and limitations \cite{xie2015performance, zhang2013restriction}. Other authors performed communication experiments exploiting optics techniques at millimeter waves \cite{yan2014high}.

In this context, we present a short range OAM based communication system implemented by means of circular arrays composed by patch antennas. Circular arrays have already been considered in literature but only as OAM generators \cite{mohammadi2010orbital, tennant2012generation}. At the best of our knowledge, this is the first experimental test in which patch arrays are used for communication purposes. 

More in detail, after this introduction, Section II reports a short theoretical frame on OAM waves and antennas properties. Section III describes the arrays designed for the multiplexing system, Section IV is dedicated to the communication tests and the obtained results while Section V examines an application of OAM waves to enhance the security level of communication systems. Finally, the last section draws up the major conclusions, underlying the most relevant aspects of this work.

\section{Background}
EM waves carrying OAM are characterized by a spiral phase front and by a doughnut-shaped intensity profile. In general, considering a cylindrical reference system $(\rho, \phi, z)$, an OAM field at $z={const.}$ can be described as $E_{\ell}(\rho, \phi)=A(\rho) \exp(i \ell \phi)$ where $A(\rho)$ is the amplitude factor, $\ell\in Z$ is the OAM value and $k$ the wave-number. 
The term $\exp(i \ell \phi)$ describes the spiral phase front of the field and denotes the presence of a central screw phase singularity \cite{berry2000making} responsible for the field central intensity minimum. Due to diffraction effect, the size of this minimum increases during propagation. For this reason, it is fundamental to use a device of suitable size to collect the whole field at a certain distance from the source.

OAM waves can be generated and detected by several devices \cite{Trinder2005,schemmel2014modular,gao2013generating,mahmouli2012orbital}. Among all, circular antennas arrays are particularly interesting because of their versatility.
For generating an OAM mode with topological charge equal to $\pm\ell$ the antennas are fed with the same signal, but then delayed relative to each other so that the phase is incremented, in a turn, by $\pm 2\pi \ell$. In other words, the overall phase between successive antennas is increased by $2 \pi |\ell|/N$, where $N$ is the number of antennas on a circle around the beam axis, in a clockwise (-$\ell$) or counter-clockwise ($+\ell$) direction.
Recall that the generation of an OAM mode of order $|\ell|$ needs an array composed by a proper number of antennas: $N > 2 |\ell| + 1$ \cite{thide2007utilization}. Otherwise, the beam phase distribution will be under-sampled, according to Nyquist theorem. 
By using circular arrays it is also possible to control the angular direction of the main lobe of the field, and the secondary lobes distribution by acting on the array radius \cite{balanis2005antenna}.
 \begin{figure*}[!ht]
\centering
 \includegraphics[width=16cm]{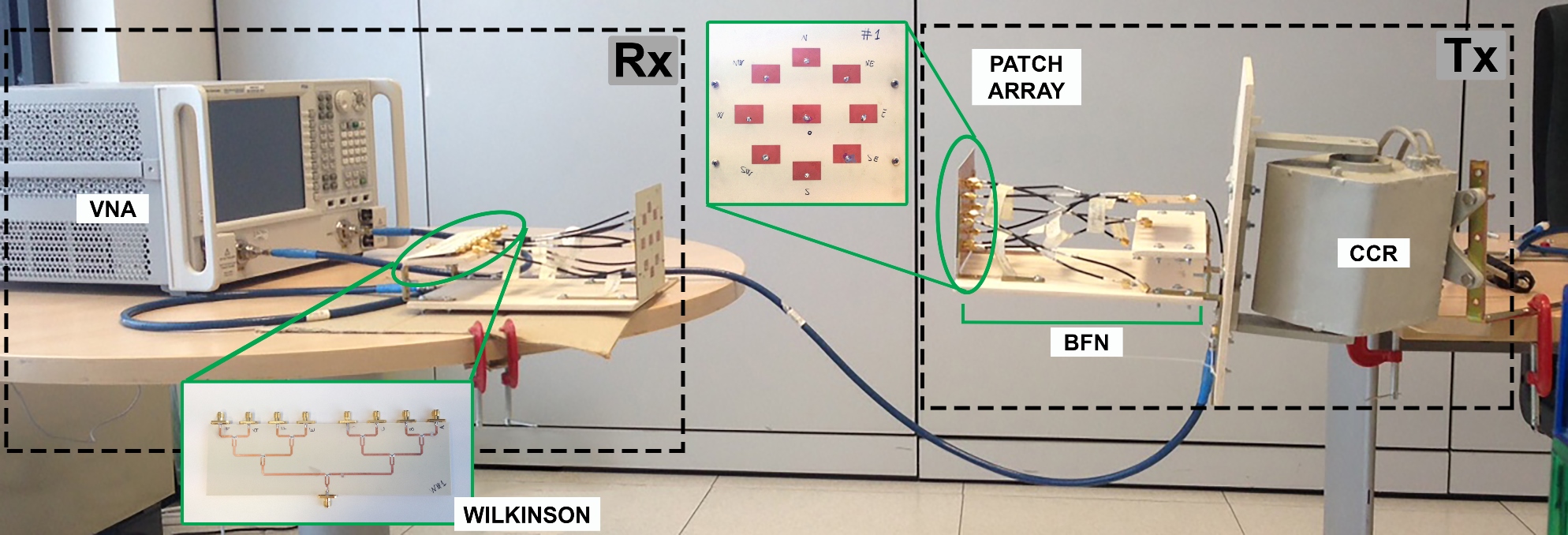}%
 \caption{Overview of the experimental setup. From left to right: Vector Network Analyzer (VNA), Rx array and Tx array with their Beam Forming Networks (BFN). Tx array is  mounted on the Computer Controlled Rotator (CCR). Left and right insets:  Wilkinson power divider and the array structure, respectively.}%
\label{fig:overview}
 \end{figure*}

Circular arrays composed by elementary dipoles \cite{thide2007utilization}, patches \cite{bai2013generation, wei2015generation} or Vivaldi antennas \cite{deng2013generation}, have been simulated for OAM generation. A patch array prototype has also been tested at the frequency of 2.5GHz \cite{bai2014experimental} but never used in a communication system.

\section{Experimental Setup}
In our experimental setup, to generate and receive OAM modes, we use two circular arrays, each composed by 9 patch antennas. 
One patch, placed at the center of the array, generates or receives the $\ell = 0$ mode, whereas eight patches evenly spaced along a circle are equipped for $\ell = \pm 1$ modes. 
Each patch is fed through a coaxial cable, linked with a SMA connector. The length of each cable is calculated in order to produce the exact phase delay of the feeding signal, defined by the desired OAM mode.
The coaxial cables are then connected to the outputs of a Wilkinson power divider, which equally splits the  input mode signal. The set of cables with the Wilkinson power divider form the so-called beam forming network (BFN). An overview of the system can be observed in Fig. \ref{fig:overview}.
To generate the $\ell = \pm 1$ modes, the coaxial cables connected to the patches of the circular array have been properly tailored, in order to induce a phase delay of $45^\circ$ between each pair of consecutive antennas. So, it is possible to induce a cumulative azimuthal phase delay to the transmitted field in a left-handed or right-handed direction, bringing to the generation of $\ell = \pm1$ modes, respectively. 

We designed and optimized the entire array, starting from single patch, of $19.58$mm width and $13.07$mm height, which has been analyzed by means of a FEM code to operate at frequency of $5.75$ GHz.
 \begin{figure}[ht]
\centering
 \includegraphics[width=8.6cm]{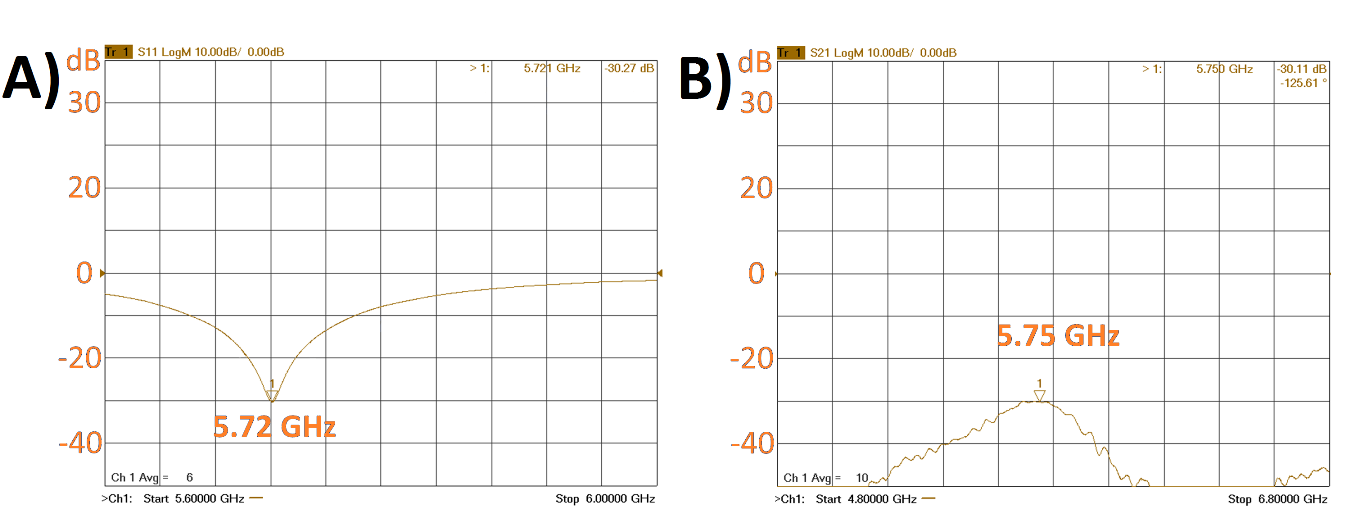}%
 \caption{A) Example of $S_{11}$ measurement of a single patch. At the resonance frequency of 5.72GHz $S_{11} = -30.3$dB. B) Example of cross-talking measurement ($S_{12}$) between the central patch of the array and the one on its left. At the working frequency of 5.72GHz $S_{12} = -30.1$dB.}%
\label{fig:RL}
 \end{figure}
Patches are realized on a ceramic copper substrate (Isola IS680-345) with a thickness of $0.75$mm and dielectric constant $\epsilon = 3.45$. The connector pin feed is centered on the non-resonant side of the patch and distant $3.57$mm from the lower edge. 
We characterized the patch antennas by using a Vector Network Analyzer (VNA), by measuring the proper $S$-parameters. 
$S_{11}$ measurements allowed to observe that the patches resonant frequency is located at $5.75$ GHz $\pm$ $30$ MHz.  
Moreover, the $S_{11}$ values at the working frequency are  below $-15$dB, thus providing a good match of the 50 $\Omega$ line impedance. A measurement example is reported in Fig. \ref{fig:RL}A.

As briefly reported, eight patches, dedicated to the generation of the $\ell = \pm 1$ OAM modes, are equally angularly disposed along a circumference with a radius of $40$mm. With this choice, the patch cross-talking value ($S_{ij}$ where $i \neq j$ are patch indexes) is predicted to be lower than $-20$dB.
Measurements confirmed that the designed array satisfies this constrain, as required. An example is reported in fig. \ref{fig:RL}B where a cross-talking diagram is shown.

A detailed list of the cables characteristic parameters are reported in Tab. \ref{tab:tab1}. First column labels the $i$th cable with $i=1,...,8$. Second and third columns contain the module and phase of the $S_{21}$ parameter, respectively, describing the input-output relation of each cable. The fourth column reports the phase delay difference between two consecutive - $i$th and $(i+1)$th - cables. Phase differences are equal to $45^\circ \pm 1^\circ$. This result is, indeed, suitable for the generation of $\ell = \pm1$ OAM modes.

\begin{table}[h]
\centering
\caption{A cable set characterization}
	\label{tab:tab1}
		\begin{tabular}{c||c|c|c}
			Cable ID & $|S_{21}|$  & $\angle S_{21}$ & $\Delta \angle S_{21}$ \\ \hline \hline
			1 & 0.41 & 117  & 44\\
			2 & 0.43 & 161  & 46\\
			3 & 0.43 & -153 & 45\\
			4 & 0.40 & -108 & 45\\
			5 & 0.42 & -63  & 44\\
			6 & 0.43 & -19  & 46\\
			7 & 0.41 & 27   & 45\\
			8 & 0.41 & 72   & 45\\ \hline
	\end{tabular}
\end{table}

As already mentioned, the second component of the BNF is the Wilkinson unit, composed by 8-line power splitter layout (see inset of Fig. \ref{fig:overview}). As first step, it has been designed and optimized by means of FEM simulations at the same working frequency of the cables, i.e., 5.75 GHz. Then, the structure has been realized on the same laminate used to built the patch. The input-output lines of the divider are $1.7$mm wide in order to realize an impedance of 50 $\Omega$; the connectors used are of SMA type. 
Again, by using the VNA, we measured the $S$-parameters. 
The resulting intensity of $S_{99}$, evaluated at the input port of the Wilkinson dividers, is lower than $-18$dB for a $100$MHz bandwidth, around the working frequency. 
Tab. \ref{tab:tab2} reports measured $S_{i9}$ parameters, with $i=1,...,8$ labeling the $i$-th divider output line.
Second and third column of the table refers to Tx divider, whereas fourth and fifth column to Rx one.
\begin{table}[h]
\centering
\caption{Characterization of two Wilkinson power divider}
	\label{tab:tab2}
		\begin{tabular}{c|||c|c|||c|c}
			OUT & $|S_{i9}|$  & $\angle S_{i9}$ & $|S_{i9}|$  & $\angle S_{i9}$\\ \hline \hline
			1 & -11.4 & 140 & -11.0 & 141\\
			2 & -11.6 & 141 & -11.1 & 140\\
			3 & -11.8 & 140 & -11.4 & 139\\
			4 & -11.5 & 142 & -11.2 & 141\\
			5 & -11.4 & 143 & -10.9 & 140\\
			6 & -11.3 & 144 & -10.9 & 141\\
			7 & -11.0 & 144 & -11.0 & 139\\
			8 & -11.0 & 143 & -11.2 & 138\\ \hline
	\end{tabular}
\end{table}
The results show that Wilkinson systems equally divides (recombine) the transmitted (received) signal both in amplitude and phase, with tolerances of $\pm 0.4$dB and $\pm 3^\circ$ respectively. 
The BFN at the Tx side was built up by connecting the $i$-th cable with the $i$-th divider output line. The same operation is made for the realization of the BFN at the Rx side.

\section{Experimental results}
We performed several experiments, using the described patch arrays. First, we generated OAM beams with $\ell=\pm1$ and then we implemented an OAM-based communication link at short range.

\subsection{Maps}
First experiments consisted in the generation of a $\ell = 1$ and $\ell = -1$ modes with our patch arrays described in the previous section. Each array, one at a time, was mounted as transmitter on a computer controlled rotator (CCR), moving along a spherical surface, both in elevation and azimuth, with a resolution of $0.5^\circ$. 
On the receiving side, a single patch antenna was placed at $0.15$m distance, in order to probe the generated field.
The input of the BFN at the Tx side was connected to the first port of a VNA, set to measure the $S_{21}$ parameter at the frequency of $5.75$GHz, and the probing patch to the second one. 
\begin{figure}[ht]
\centering
 \includegraphics[width=9cm]{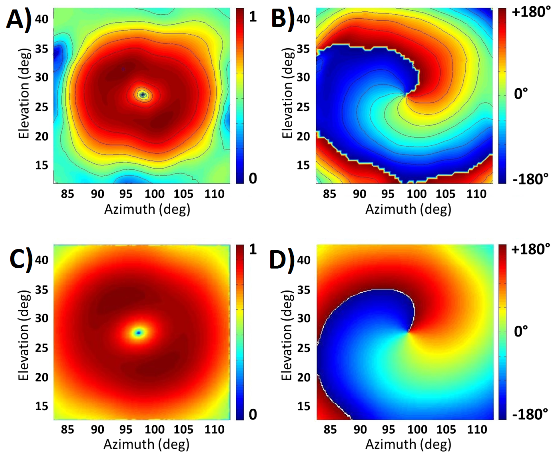}%
 \caption{Electric field generated by a 8 patches circular array configured to produce a $\ell = +1$ OAM beam. Measurement (A) and simulation (C) of the normalized intensity distribution. Measurement (B) and simulation (D) of the phase distribution.}%
\label{fig:l+1}
\end{figure}
\begin{figure}[ht]
\centering
 \includegraphics[width=9.05cm]{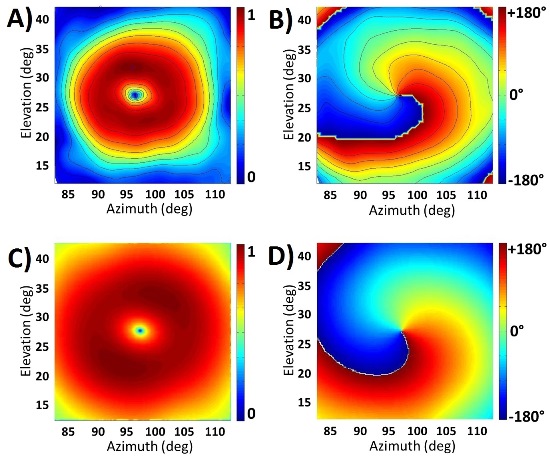}%
\caption{Electric field generated by a 8 patches circular array configured to produce a $\ell = -1$ OAM beam. Measurement (A) and simulation (C) of the normalized intensity distribution. Measurement (B) and simulation (D) of the phase distribution.}%
\label{fig:l-1}
\end{figure}
The VNA was set to transmit a dummy sinusoidal signal of $0$dBm power. All the measurements were performed in free space and collected into 2D maps. 
The experimental results have been compared with numerical FEM based simulations.
Two examples of simulated and measured maps, of $\ell = 1$ and a $\ell =-1$ beam, are reported in Fig. \ref{fig:l+1} and \ref{fig:l-1}, respectively. 

Azimuth and elevation values reported on the map axis correspond to the coordinates of the CCR on which the Tx array is mounted. So, the center of the map corresponds to the configuration in which the transmitting and receiving antennas are aligned.

A good match between the expected fields and the measured ones is found. In fact, the experimental fields are characterized by a well defined doughnut distribution and the beam phase presents the characteristic vortex shape with the central singularity.  Vortices turn in clockwise or counterclockwise direction, according to the negative or positive OAM value, by convention. 
We calculated, by applying Friis equation \cite{someda2006electromagnetic} to experimental results, a maximum gain of about $10.5$ dBi both for $\ell = 1$ and a $\ell =-1$ modes radiation patterns, in good agreement with FEM simulations prediction of about $11$dBi.

\begin{figure}[ht]
\centering
 \includegraphics[width=8.8cm]{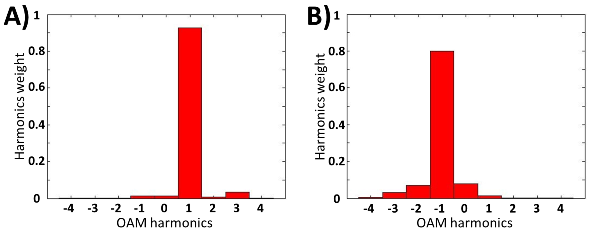}%
 \caption{Spiral spectrum decompositions of the OAM field generated by circular arrays. (A) spectrum of the $\ell = +1$ field shown in Fig. \ref{fig:l+1}; (B) spectrum of the $\ell = -1$ field shown in Fig. \ref{fig:l-1}. The fundamental harmonic, in both cases, carries $90\%$ and $85\%$ of total fields energy, respectively.}%
\label{fig:spiral_spectrum}
\end{figure}

We have also calculated the OAM content carried by the experimental fields. For this purpose, we applied the spiral spectrum algorithm \cite{torner2005digital} which consists in the projection of the EM field on spiral harmonics, ($\exp{(i \ell \phi)}$ terms), similarly to a Fourier transform. The spectra of the measured fields are shown in Fig. \ref{fig:spiral_spectrum}. 

As reported, the fundamental azimuthal harmonics carry $90\%$ and $85\%$ of total energy in correspondence of the $\ell = 1$ and $\ell = -1$ mode, respectively. Small field imprecisions generate the spurious OAM modes reported in Fig. \ref{fig:spiral_spectrum}.
This spectra confirm the good quality of the generated OAM modes.

\subsection{OAM modes synthesis robustness}
The second series of experiments tested the dependence of the generated OAM fields on the number of radiating patch antennas.
First of all, we sampled and observed an $\ell=1$ field generated by eight radiating elements. In subsequent experimental steps the patch antennas were progressively disconnected one at a time up to only one connected antenna.
Patch antennas were switched off by disconnecting the feeding coaxial cables. In order to prevent impedance mismatch at the Wilkinson output, the corresponding coaxial cables were connected to a $50 \Omega$ dummy load.

\begin{figure}[ht]
\centering
 \includegraphics[width=8.9cm]{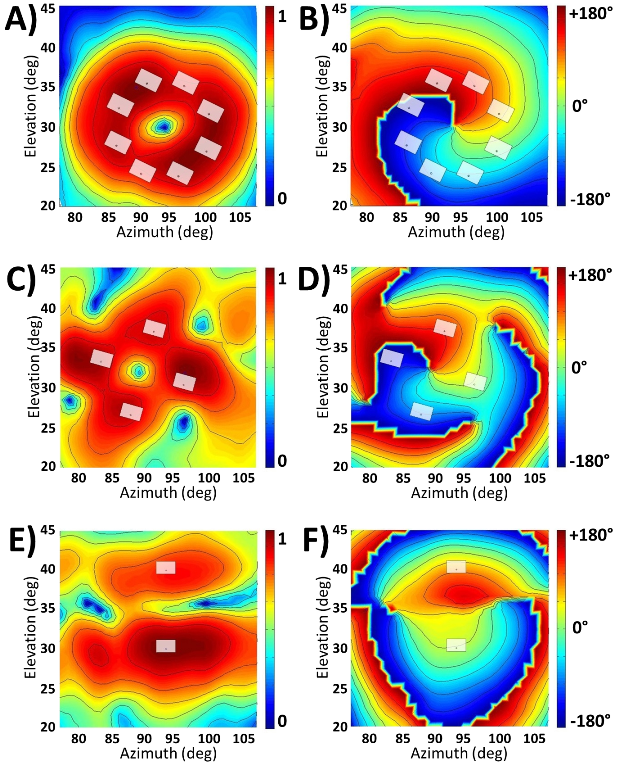}%
 \caption{Normalized intensity (A), (C), (E) and phase (B), (D), (F) distributions of a $\ell = +1$ OAM beam generated by arrays composed by 8, 4 and 2 patches, respectively. The patches position is represented, both on amplitude and phase maps, with superimposed gray rectangles.}%
\label{fig:patch_lack}
\end{figure}

Fig. \ref{fig:patch_lack} reports some meaningful examples of the field distributions, both in amplitude and phase, generated by array configurations with an even number of patches. 

The arrangement of the active patches has been superimposed to each figure of the amplitude field distribution. 
The most important aspect to be stressed is that the phase singularity is present only when the sampling theorem is satisfied. In fact, the antennas number, $N_a$, has to satisfy the condition $N_{a} > 2 |\ell| + 1$, which corresponds to $N_{a} > 3$ for generating an $\ell = + 1$ mode. For this reason in Fig. \ref{fig:patch_lack}D, corresponding to the under sampled case $N_a =2$,  the phase singularity is not longer present. So, when $N_a < 2$,  the number of active elements is not enough to reproduce neither intensity nor phase profile of an $\ell =1$ OAM mode.

\subsection{OAM Communication}
Here we report the experimental results of a communication system based on the mutual orthogonality of OAM states.

We briefly recall, in terms of geometrical properties, how the orthogonality plays a role in the OAM communication.
When an OAM antenna, (in our case an array and the corresponding BFN), is used in the reception side, it behaves as an inverse phase antenna, because of the change of propagation direction \cite{mari2015near}.  
In particular, our Rx OAM antenna imparts to the received mode, an azimuthal phase delay $\exp(\mp i|\ell|\phi_i)$, being $\phi_i$ the angular position of the $i$-th patch element with $i=1,...,8$, opposite to that it would impart if used as in transmission. Since the normal vector of the Rx array plane is opposite to the propagation axis of Tx one, an incoming field with left-handed (right-handed) topological charge is sampled in a right-handed (left-handed) way by the Rx array elements. In other words, a transmitting $\ell= +1$ antenna receives like a $\ell = -1$ one. As a consequence, the output field of the receiving OAM antenna is characterized by the topological charge of the impinging beam, plus the topological charge of the antenna in reception mode.

In order to test an OAM-based communication, a system has been assembled with the same setup used for producing the maps presented in the previous sections. The Tx array was mounted on the CCR while the Rx one was fixed at a distance of 0.5m from Tx. 
The VNA, connected to the input and output ports of the BFN at Tx and Rx side, respectively, measured $S_{21}$ parameter, both in amplitude and phase.
We evaluated the channel gain for each position of the Tx array. Resulting data are reported on 2D maps.

In the first test, a transmitting $\ell=+1$ antenna communicates with an identical receiving one ($\ell= - 1$ in reception mode, according to the change of propagation direction). In other words, the Tx array was pre-coded by means of its BFN to produce a $\ell = +1$  mode while the Rx antenna acted in order to impart to the received field an opposite azimuthal phase delay $\exp(- i|\ell|\phi_i)$. In this way, the BFN at the receiving side acts destructively on the azimuthal phase carried by the transmitted field and reset to zero the transmitted topological charge from $\ell = +1$ to $\ell=0$. An analogous behavior presents a $\ell = -1$ antenna that acts in pair with an identical $\ell = -1$ one

\begin{figure}[ht]
\centering
 \includegraphics[width=8.9cm]{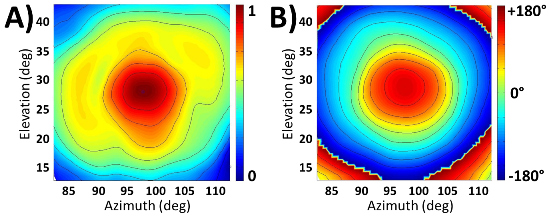}%
\caption{Normalized intensity (A) and phase (B) distributions of the electric field produced by a $\ell = +1$ wave received by means of a $\ell = -1$ circular array. 
As expected, the topological charge of the $\ell = +1$ wave is reset to zero by the receiving array: $\ell_{\text{wave}} + \ell_{\text{Rx}} = +1 -1 = 0$.}
\label{fig:tx1rx1}
\end{figure}

\begin{figure}[ht]
\centering
 \includegraphics[width=9cm]{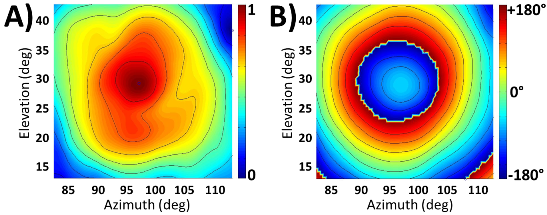}%
 \caption{Normalized intensity (A) and phase (B) distributions of the electric field produced by a $\ell = -1$ wave received by means of a $\ell = +1$ circular array.  
As expected, the topological charge of the $\ell = -1$ wave is reset to zero by the receiving array: $\ell_{\text{wave}} + \ell_{\text{Rx}} = -1 +1 = 0$.}
\label{fig:tx-1rx-1}%
 \end{figure}

Both intensity and phase distribution of Fig. \ref{fig:tx1rx1} and Fig. \ref{fig:tx-1rx-1} resembles a standard $\ell=0$. 
These results prove that two array of patch antennas designed for the same OAM topological charge can efficiently communicate one to each other.

\begin{figure}[ht]
\centering
 \includegraphics[width=8.9cm]{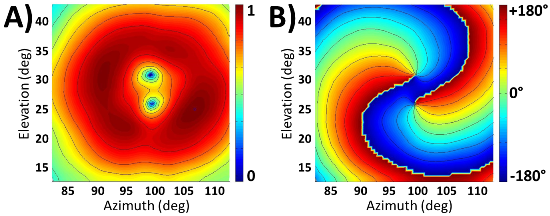}%
 \caption{Normalized intensity (A) and phase (B) distributions of the electric field produced by a $\ell = 1$ wave received by a $\ell = +1$ circular array. 
An increase of the topological charge of the transmitted field is observed ($\ell_{\text{wave}} + \ell_{\text{Rx}} = +1 +1 = +2$).}
\label{fig:tx1rx-1}%
 \end{figure}

As second test, the Tx array set to produce a $\ell=+1$ operates in pair with an opposite one, i.e., a patch array with BFN set to produce a $\ell=-1$ that acts in reception mode as a $\ell=+1$, as stated above.
The behavior of the communication system is significantly different, because the received twisted field is transformed into an $\ell = + 2$ beam, (see Fig. \ref{fig:tx1rx-1}).

In this case, based on the action of the BFN at the Rx side, the resulting azimuthal phase delay induced to the transmitting mode is doubled. Hence, the distribution, both in intensity and phase, can be attributed to an $\ell = 2$ mode. 
The split of singularities, visible in Fig. \ref{fig:tx1rx-1}, is due to small imprecision in the experimental setup, as observed in a vast literature when generating high order OAM modes \cite{ricci2012instability}. 
In our experimental setup, deviations are mainly caused by inaccuracies of the arrays and uncertainties on the feeding lines of the patch antennas. In fact, the feeding signals are affected by little errors in amplitude and phase, as reported in Tab. \ref{tab:tab1}-\ref{tab:tab2} of the previous section. Even if these errors are very small, the field patterns generated by Tx array are not perfectly balanced. Moreover, the signal received by the patch antennas of Rx array is not equally recombined.

\subsection{Tolerance with respect to ideal position}
The fourth series of experiments shows how the OAM-based communication is sensitive to lateral shift and to angular tilt between the plane of the transmitting array and that of the receiving one. 
\begin{figure}[ht]
\centering
 \includegraphics[width=8.5cm]{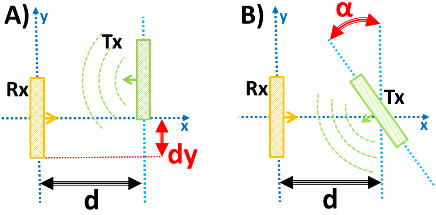}%
 \caption{Outline the Tx array movements: (A) translation and (B) tilt. $d$ represents the distance between Tx and Rx arrays while $d_y$ and $\alpha$ represent the translation and tilt offsets, respectively.}
\label{fig:movements}%
 \end{figure}
For this purpose, we developed two experimental tests.
First, we tested the system tolerance to a lateral shift of the transmitting array, (see Fig. \ref{fig:movements}A). We measured the power of $\ell=0$ and $\ell=1$ modes, transmitted one a at time at different shifted position with respect to the perfect alignment, and received by a $\ell=0$ antenna.
The $\ell=1$ mode was generated by 8 patch antennas, pre-processed by means of a proper BFN, while the central single patch antenna generated the standard $\ell=0$ beam. 
Both the $\ell=0$ and $\ell=1$ modes were fed and piloted by a signal switch connected, in turn, to the first port of the VNA. 

\begin{figure}[ht]
\centering
 \includegraphics[width=8cm]{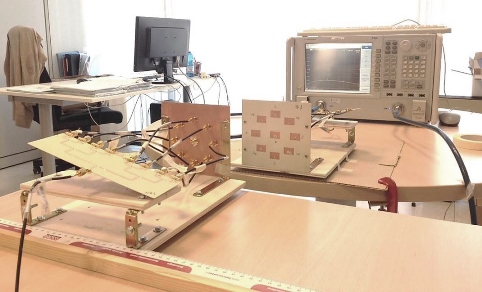}%
\caption{Experimental setup for the evaluation of the tolerance with respect to a lateral shift of the Tx array. The ruler, used to measure the translation shift is visible on the bottom side.}
\label{fig:shift_setup}%
 \end{figure}

\begin{figure}[ht]
\centering
 \includegraphics[width=8.3cm]{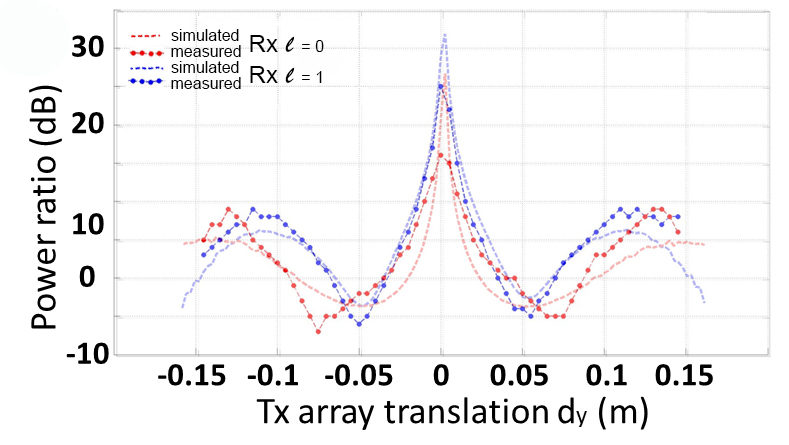}%
 \caption{Measured and simulated power ratio as function of translated position $d_y$. Blue lines:  power ratio between $\ell=1$ and $\ell=0$ transmitted modes when received by an $\ell = 1$ antenna.
Red lines: power ratio between $\ell=0$ and $\ell=1$ transmitted modes when received by a $\ell = 0$ antenna.
}
\label{fig:snir_tr}%
 \end{figure}

These two modes were received by a single patch antenna, connected to the second port of the VNA. The instrument was set to measure the $S_{21}$ parameter. Tx and Rx antennas have been positioned face to face at 0.15m distance. Rx array was kept fixed, while the Tx one was bound to translate along a ruler parallel to the Rx array (see Fig. \ref{fig:shift_setup}).
At each translation step we transmitted, by acting on the switch at Tx side, once the $\ell =0$ mode once the $\ell = +1$ one.
In this way, the receiving single antenna probed powers alternately coming from the two transmitted modes. We calculated the power ratio between an $\ell=0$ and $\ell=1$ modes, i.e., the power ratio between the mode identical to the one used in reception and the one with different OAM value. Fig. \ref{fig:snir_tr}A (red line) reports the power ratio between $\ell=0$ and $\ell=1$ as function of lateral shift.

The entire procedure was repeated replacing the receiving single antenna with an $\ell=1$ array, which imparts an azimuthal $\exp(-i\ell\phi_i)$ phase delay to the received modes.
As a consequence, in this case, we calculated the power ratio between an $\ell=1$ and an $\ell=0$ mode. Fig. \ref{fig:snir_tr}A (blue line) reports this value for different positions.
Both lines in Fig. \ref{fig:snir_tr}A are characterized by a single central maximum that corresponds to the perfect alignment between Tx and Rx arrays. This confirms the importance of the alignment for an optimal modal isolation. 
The curve maximum of the $\ell = 1$ channel is higher than the $\ell = 0$ one. This is caused by the different number of active antennas (8 for the $\ell=1$ mode and one for the $\ell=0$ mode).
Both lines present steep edges: a little translation, with respect to the alignment position, is sufficient to cause high degradation of the power ratio. In particular, in our experimental conditions, it is sufficient $1$cm shift to significantly enhance a modal interference with a drop of about $10$dB in the power ratio.

For the sake of completeness, we have also simulated these communication experiments by means of a numerical MIMO based model \cite{hampton2013introduction} (see the Appendix). Simulations have been performed taking into account the 
amplitude and phase errors on the patch feeding signals, reported on Tab. \ref{tab:tab1}-\ref{tab:tab2}.
The simulated results are reported in Fig. \ref{fig:snir_tr}B.
A good match between the experimental results and the simulations is seen. In particular, curves behavior, positions of each maximum, minimum and inflection points are coincident with a good approximation. On the contrary, a little discrepancy about the peaks values is shown. This deviation can be mostly explained by the mathematical model applied for the simulations. In fact, the simulated arrays are composed by elementary point-like antennas and by loss-free BFN. As a consequence, the energy dissipation of a realistic BFN and the coupling between patches and EM fields are not considered.
Another contribution to the discrepancy is given by the manual translation of the arrays, which leaded to a position uncertainty of about $0.2$mm.

\begin{figure}[ht]
\centering
 \includegraphics[width=7cm]{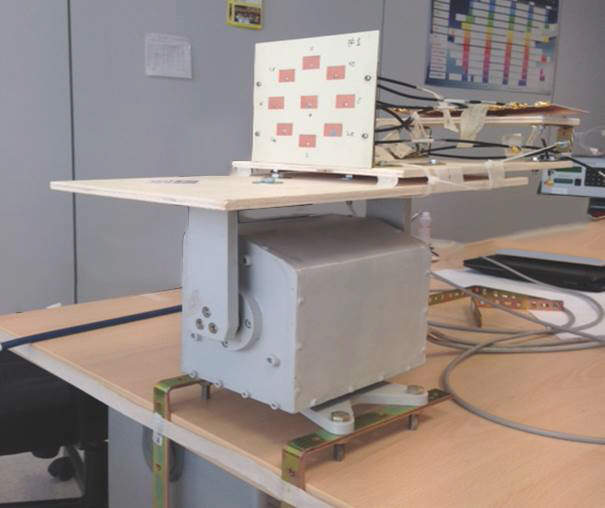}%
 \caption{Experimental setup for the evaluation of the tolerance with respect to a angular tilt of the Tx array. Tx array is fixed on a CCR. These two devices share the same rotation axis.}
\label{fig:setup_tilt}
 \end{figure}

\begin{figure}[ht]
\centering
 \includegraphics[width=8.3cm]{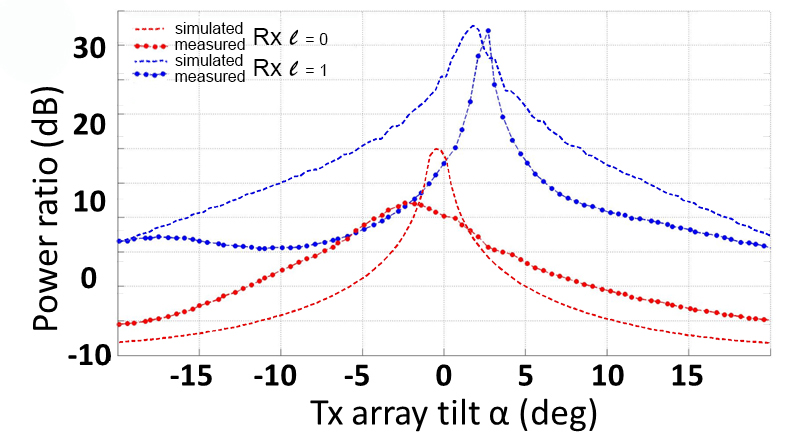}%
 \caption{
Measured and simulated power ratio as function of angularly tilted position $d_y$. Blue lines:  power ratio between $\ell=1$ and $\ell=0$ transmitted modes when received by an $\ell = 1$ antenna.
Red lines: power ratio between $\ell=0$ and $\ell=1$ transmitted modes when received by a $\ell = 0$ antenna.
}
\label{fig:snir_tl}%
 \end{figure}

A second set of experimental tests was performed in order to evaluate the sensitivity of an OAM-based communication system to the angular tilt between Tx and Rx arrays, (see Fig. \ref{fig:movements}B).
In Fig. \ref{fig:setup_tilt} we show the Tx antenna, connected to the switch and fixed to the CCR. The antenna transmitted once the $\ell =0$ mode and alternately the $l=+1$ mode only.
We calculated again the power ratio between the mode identical to that used in reception and the one with different value of OAM, for the receiving antenna set once to $\ell=1$ and once to $\ell=0$. These values were collected in function of the tilting angle of  the Tx array. 
Fig. \ref{fig:snir_tl} shows the experimental and simulated results.
The red and blue lines indicates that the Rx antenna is a single patch or a circular array, respectively. Again, the central peak corresponds to the alignment condition.
A tilting angle of $5^\circ$ is enough to obtain a drop of about $10$dB in the power ratio for both lines.
A tilting angle of $5^\circ$ is enough to obtain a drop of about $10$dB in the power ratio for both lines.

We can observe existing similarities and some differences between the simulated and the measured power ratios. The main differences concern the curves slope, especially near the peaks and their maximum values.
The main differences observed between the numerical simulations and the corresponding experimental results are due to a non perfect alignment between the axis of the CCR and the one of the Tx array, since it was manually placed. Moreover, the mechanical limit of our setup could have caused further small misalignment during the measurements. 
Another contribution to the mismatches is given by the non-negligible error on the CCR positioning, equal to $0.2^\circ$ according to technical data sheet.
As in the previous case, the simulations have been performed considering arrays composed by elementary point-like antennas, fed by loss-less BFN. So, once again, a lower power level of experimental curves is observed.

We can also notice that the peaks of the experimental power ratio are in different angular positions. This behavior can be attributed to amplitude and phase errors introduced by the BFN, as confirmed by simulated curved, obtained by introducing small random deviations in the matrices describing the BFN behavior.
In conclusion, we have shown, both with simulations and experimental data, that in a communication system based on OAM modes the alignment between Tx and Rx antennas is of paramount importance. Moreover, special attention has to be paid also to the BFN of Tx and Rx antennas in order to avoid unwanted inaccuracies.

\section{OAM communication: an application example}
The results presented in the previous section show how the relative position between the Rx and Tx array deeply affect the communication efficiency of a Line Of Sight (LOS) OAM-based link. 
In order to fully exploit OAM mode orthogonality, the Tx and Rx array planes should be aligned (Fig. \ref{fig:snir_tr}, \ref{fig:snir_tl}) and at proper distance, such that the Rx array efficiently couples to the main lobe (doughnut-shaped) of the emitted radiation. Deviations from this optimal configuration, in general, deteriorates the communication performances.
These geometrical constraints limit the feasibility of an OAM-based communication,
since the use of OAM modes restricts the physical space suitable for receiving.
On the other hand, this characteristic can be of special interest for applications in which it is of crucial importance that the information exchanged between Tx and Rx must not be intercepted by third parties, like the exchange of personal data, electronic payments, etc.

We can compare an OAM-based communication link with respect to a standard one, showing that the physical space in which one can be intercepted is much smaller by using OAM modes.
To better explain this concept, we consider an example of communication link under the following hypothesis: 
\begin{itemize}
\item{Tx and Rx antennas are circular arrays comprising 9 patch antennas linearly polarized. As described in Section III, one patch is equipped for the $\ell = 0$ mode and 8 patches, placed along a circle, for the $\ell = +1$ one;} 
\item{the mutual coupling between patches of the same array is neglected;}
\item{system operative frequency is equal to $f_0=5.75$GHz;}
\item{the radius of each array is equal to $R=0.05$m;}
\item{BFNs of both arrays are loss-free and designed to transmit, or receive, three signals associated to the $\ell =-1$, $\ell = 0$ and $\ell =+1$ OAM modes;}
\item{the three signals are transmitted with the same power;}
\item{Tx and Rx arrays are normally faced and aligned at a distance $d = 0.15$m, (note that $d$ is smaller than the arrays Fraunhofer length: $d_F = 2R^2/\lambda = 0.87$m thus allowing OAM channel multiplexing \cite{edfors2012orbital}). The quantity $d$ has been calculated applying the mathematical model described in Appendix and imposing, for the three channels, a SNIR equal to $25$dB;}
\item{each of the three channels supports a digital modulation whose bit error rate is negligible (i.e. below a given threshold for which the chosen error-correction code practically yields an error-free communication) as long as the channel SNIR is larger than $15$dB;}
\end{itemize}
As mentioned in the previous section, a misalignment between Tx and Rx antennas causes a degradation of OAM modes orthogonality and consequently an enhancement of channels cross-talking. To quantify this effect, the mathematical model described in the Appendix can be adopted to calculate the channels SNIR at different positions of the Rx array, while keeping fixed the Tx one. Through numerical simulation it is possible to determine the spatial bounds within which the three channels can be received without errors.

The first simulation studies the effects on the SNIR of a parallel shift between the Rx array center and the Tx one, (similarly to Fig. \ref{fig:movements}A). Each position of the Rx array is identified by two parameters: the lateral displacement $d_y$ and the distance $d$ with respect to the Tx array.
Based on the results a receiving map can be calculated. 
\begin{figure}[ht]
\centering
 \includegraphics[width=9cm]{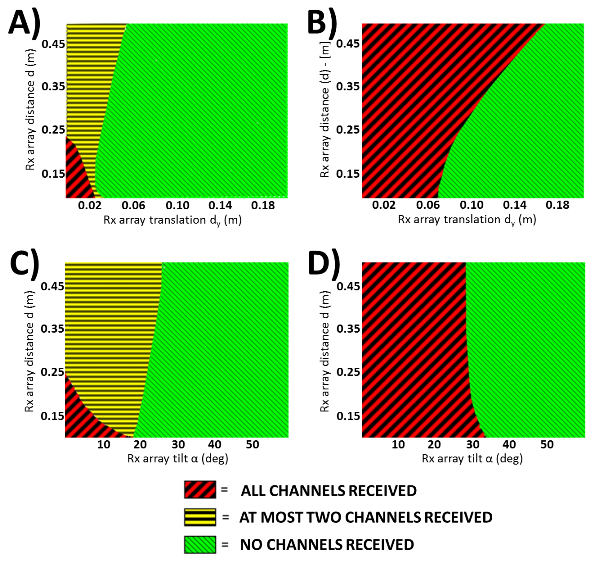}%
 \caption{Maps of reception zones for multi-channels OAM based communication system (A), (C)  and for single channel traditional ones (B), (D). Vertical axis reports the distance between Tx and Rx arrays. Horizontal axis indicates lateral shift (A), (B) or angular tilt (C), (D) between Rx and Tx array. Red areas indicates physical space where the Rx array receives three channels. In yellow and green areas at most two channels can be correctly received.}
\label{fig:map}%
 \end{figure}
As can be observed in Fig. \ref{fig:map}A, the three channels can be received only in a restricted area, identified by red lines, near the alignment position. On the contrary, in the yellow zone and in the green one, at most two channels can be correctly received. As previously suggested, this particular behavior can be exploited to enhance the communication security: if the transmitted information is split over the three channels, it is more difficult for an eavesdropper to catch and reconstruct the whole data stream exchanged between Tx and Rx. The only way for an eavesdropper to intercept the information is to stay very close and almost aligned to the legitimate Rx.

To better appreciate the security advantage of an OAM-base communication system, it is useful to repeat the calculation with a different configuration of the communication system. In this case, it has the same antenna structure but it is based on the transmission of a single channel over a traditional $\ell = 0$ mode. The assumptions about channel modulation properties are the same. This new configuration allows to simulate the behavior of a traditional communication system and to compare it with the previous OAM-based one. The results obtained from simulations are collected in the map of Fig. \ref{fig:map}B. As can be noticed, this second map is composed only by two zones since only a single channel is involved in the communication process. Moreover, the red zone where the information can be received is much larger. By comparing the two maps of Fig. \ref{fig:map}A and B it is clear that the second configuration is more vulnerable to eavesdropping because of the larger size of the reception area.

To complete the study, it is interesting to repeat the two aforementioned simulations in presence of an angular tilt of the Rx array. In particular, each position of the Rx array is now identified by a tilt angle $\alpha$ and by the distance $d$ with respect to the Tx array, (similarly to Fig. \ref{fig:movements}B). All the results are collected in the maps of Fig. \ref{fig:map}C and Fig. \ref{fig:map}D, for the OAM-based system and for the traditional one, respectively. 
As expected, also in this second configuration the OAM based system is characterized by a slightly smaller reception area respect to the traditional one. In fact, the tilt misalignment prevent the Rx array to correctly recognize the transmitted OAM modes with a consequent increase of channels cross-talking. 

By comparing the results of Fig. \ref{fig:map}A-B with the ones of Fig. \ref{fig:map}C-D, it is clear that multi-channel systems can be smartly employed to enhance, the security level of traditional communication systems. To implement such solutions, different set of orthogonal radiation modes may be considered. However, among all, OAM beams are  of particular interest. In fact, they are characterized by a regular and symmetric distribution of the EM field which can be easily predicted \cite{parisi2014manipulating} in a symmetric cylindrical reference system. Moreover, the OAM modes are naturally orthogonal but exploiting this property in a communication link strongly relies on antenna setup. As a consequence, a simple misalignment destroys modes Rx orthogonality, thus enhancing channel cross-talking and preventing the reception from an unauthorized user. Finally, an OAM-based multiplexing system requires fixed antenna structures and do not need further digital post processing. For these reasons, the use of OAM modes in the near field range should not be limited only to the increasing of communication capacity \cite{huang2014100} but it should be taken under consideration also to enhance the communication security in a simple and practical ways.

\section{Conclusions}
In this work we have studied and evaluated short range LOS communication systems based on OAM-modes multiplexing. We designed two circular arrays, composed by patch antennas, in order to generate and detect OAM modes with $\ell =-1, 0, +1$ value, at the frequency of $5.75$GHz. 
A very good agreement with theoretical predictions is found. Hence, the arrays are employed as Tx and Rx antennas to implement a short range radio link based on $\ell =-1, 0, +1$ modes. 
Numerical and experimental investigations have shown that when TX and RX OAM antennas are aligned channels exhibit high reciprocal isolation, thanks to the intrinsic modal orthogonality. This confirms and fosters the idea of OAM as means for carrying high-speed information over multi-channel links with low-complexity processing \cite{bozinovic2013terabit}.
Moreover, based on these results, we propose an OAM-based systems to implement secure communication links that may prevent the fraudulent interception of information by eavesdroppers or third parties. The potential of this new application is confirmed also by numerical simulations in which the security enhancement of OAM-based systems is clearly underlined.
In conclusion, the results presented in this work confirm that OAM modes are powerful tools not only to improve communication systems capacity with low complexity, but also to enhance their security down at the physical level, notwithstanding any higher-level security protocol (cryptographic keys, authorisation systems...).

\begin{figure}[ht]
\centering
 \includegraphics[width=8cm]{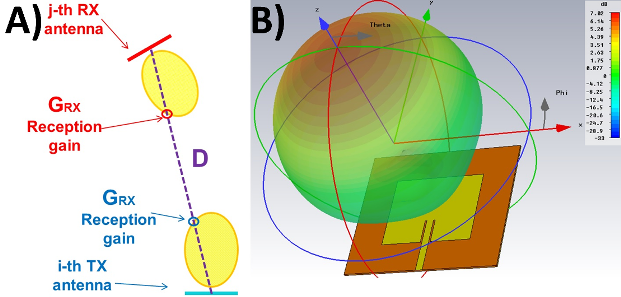}%
 \caption{A: calculation of the channel matrix ${\bf H}$ accounting for the antennas directivity. B: example of a patch radiation pattern calculated by means of a FEM simulation.}%
\label{fig:directivity}
 \end{figure}

\appendix{} 
This appendix describes the mathematical model, based on Multiple Input Multiple Output (MIMO) formalism \cite{edfors2012orbital, hampton2013introduction}, used to simulate the behavior of the communication systems previously described.\\
The mathematical model of a simple MIMO system can be expressed by means of linear algebra. Input and output data streams are mathematically described with column vectors called $\mathbf x $ and $\mathbf y$, respectively. The beam forming networks 
are described by two matrices: ${\mathbf P}$ for the Tx array and ${\mathbf Q}$ for the Rx one. ${\mathbf P}$ and ${\mathbf Q}$ are also be called precoder and postcoder matrices respectively. The communication channel is described by a matrix ${\mathbf H}$ whose entries characterize the relationship between each couple of antennas belonging to the Tx and Rx arrays respectively. 
All these elements are related to each other by the equation:
\begin{equation}
{\bf y}={\bf Q}^{\dag} {\bf H} {\bf P} {\bf x} + {\bf y_n}
\label{eq:y}
\end{equation}
where ${\mathbf y_n}$ is the white noise at the receiver and ${\dag}$ stands for complex conjugate operator. \\
The communication systems examined in this work are composed by circular arrays, in which each data stream is transmitted over a different OAM mode. For this reason, the entries of  rows and columns of pre- and post-coder matrices are equal to:
\begin{equation}
P_{n, \ell} = \frac{1}{\sqrt{N}}\exp \left( -i 2 \pi \ell \frac{n}{N} \right)
\label{eq2},
\end{equation}
\begin{equation}
Q^{\dag}_{\ell^{'},s} = \frac{1}{\sqrt{N}}\exp \left( +i 2 \pi s \frac{\ell^{'}}{M} \right)
\label{eq22},
\end{equation}
where $\ell$ and $\ell^{'}$ are the OAM modes to be transmitted and received, respectively; $N$ ($M$) is the number of transmitting (receiving) antennas and $n \in \{0,...,N-1\}$, $s \in \{0,...,M-1\}$. According to our experimental configuration, the $\ell = 0$ mode is transmitted by a central antenna. Moreover, the model has been refined taking into account, within the ray tracing method, the directivity of the antennas. In such a way, the entries of ${\bf H}$ have been calculated as:
\begin{equation}
h_{m,n}= D \frac{\lambda}{4 \pi r_{m,n}}\sqrt{G_{m,n}^{(T)}}\sqrt{G_{m,n}^{(R)}}e^{-ik r_{m,n}}
\label{eq:eq5}
\end{equation}
where $r_{s,n}$ is the geometric distance between each antenna element while weights $G_{s,n}^{(T)}$ and $G_{s,n}^{(R)}$ are the transmission and reception gain respectively, taken from the FEM simulation data, see Fig. \ref{fig:directivity}. $D$ is the center-to-center distance between the two arrays and has been introduced in order to normalize the MIMO free space  loss coefficient, $L_{FS}^{MIMO}=4\pi r_{s,n}/\lambda$, with that of a single-input single-output (SISO) system, $L_{FS}^{SISO}=4\pi D/\lambda$. The SISO transmitted power is equal to $P_{SISO}$.
Using the model just described it is quite simple to determine the MIMO Signal to Noise and Interference Ratio (SNIR) of each received data stream. In particular, the SNIR normalized with respect to the Signal to Noise Ratio (SNR) of a reference SISO link over the same distance and with the same power and noise per channel. Within a MIMO system, the SNIR of a data stream is defined as:
\begin{equation} 
\text{SNIR}_{dB} = 10\log_{10}\frac{P_s}{(P_i+P_n)}
\label{eq:eq8}
\end{equation}
where $P_i$ is the stream received power while $P_i$ and $P_n$ are the interference and noise powers respectively. Now, letting each stream to be transmitted with a power equal to $P_{\text{SISO}}$ it is possible to express Eq. \ref{eq:eq8} by means of Eq. \ref{eq:eq5}. It results that: $P_s = k_{ii}^2 P_{\text{SISO}}$, $P_i = \sum_{j \neq i} k_{ij}^2 P_{\text{SISO}}$ and $P_n = \sigma^2 \sum_{j} q_{ij}^2 $ where the $k_{ij}$ and $q_{ij}$ coefficients come from the matrices ${\bf K} = {\bf Q}{\bf H} {\bf P}$ and ${\bf Q}$. Finally, remembering that for the SISO reference link the SNR is equal to $\text{SNR} = P_{\text{SISO}}/ \sigma^2$ it results that
\begin{equation}
\text{SNIR}_{dB} = 10 \log_{10} \Biggr( \frac{k_{ii}^2}{\text{SNR}\sum_{j \neq i} k_{ij}^2 + \sum_{j} q_{ij}^2} \Biggr)
\label{eq:snir}
\end{equation}
%

\section*{Acknowledgment}
The authors acknowledge the logistic and financial support of SIAE Microelectronics in the designing, building, and testing of the setup.

\ifCLASSOPTIONcaptionsoff
  \newpage
\fi


%
%


\begin{thebibliography}{10}

\bibitem{allen1992orbital}
Les Allen, Marco~W Beijersbergen, RJC Spreeuw, and JP~Woerdman.
\newblock Orbital angular momentum of light and the transformation of
  laguerre-gaussian laser modes.
\newblock {\em Physical Review A}, 45(11):8185, 1992.

\bibitem{yao2011orbital}
Alison~M Yao and Miles~J Padgett.
\newblock Orbital angular momentum: origins, behavior and applications.
\newblock {\em Advances in Optics and Photonics}, 3(2):161--204, 2011.

\bibitem{furhapter2005spiral}
Severin F{\"u}rhapter, Alexander Jesacher, Stefan Bernet, and Monika
  Ritsch-Marte.
\newblock Spiral phase contrast imaging in microscopy.
\newblock {\em Optics Express}, 13(3):689--694, 2005.

\bibitem{mari2012sub}
E~Mari, F~Tamburini, GA~Swartzlander, A~Bianchini, C~Barbieri, F~Romanato, and
  Bo~Thid{\'e}.
\newblock Sub-rayleigh optical vortex coronagraphy.
\newblock {\em Optics express}, 20(3):2445--2451, 2012.

\bibitem{wang2012terabit}
Jian Wang, Jeng-Yuan Yang, Irfan~M Fazal, Nisar Ahmed, Yan Yan, Hao Huang,
  Yongxiong Ren, Yang Yue, Samuel Dolinar, Moshe Tur, et~al.
\newblock Terabit free-space data transmission employing orbital angular
  momentum multiplexing.
\newblock {\em Nature Photonics}, 6(7):488--496, 2012.

\bibitem{thide2007utilization}
Bo~Thid{\'e}, H~Then, J~Sj{\"o}holm, K~Palmer, Jan Bergman, TD~Carozzi, Ya~N
  Istomin, NH~Ibragimov, and Raisa Khamitova.
\newblock Utilization of photon orbital angular momentum in the low-frequency
  radio domain.
\newblock {\em Physical review letters}, 99(8):087701, 2007.

\bibitem{huang2014100}
Hao Huang, Guodong Xie, Yan Yan, Nisar Ahmed, Yongxiong Ren, Yang Yue, Dvora
  Rogawski, Moshe~J Willner, Baris~I Erkmen, Kevin~M Birnbaum, et~al.
\newblock 100 tbit/s free-space data link enabled by three-dimensional
  multiplexing of orbital angular momentum, polarization, and wavelength.
\newblock {\em Optics letters}, 39(2):197--200, 2014.

\bibitem{mari2015near}
Elettra Mari, Fabio Spinello, Matteo Oldoni, Roberto Ravanelli, Filippo
  Romanato, Giuseppe Parisi, et~al.
\newblock Near-field experimental verification of separation of oam channels.
\newblock {\em Antennas and Wireless Propagation Letters, IEEE}, 14:556--558,
  2015.

\bibitem{tamburini2012encoding}
Fabrizio Tamburini, Elettra Mari, Anna Sponselli, Bo~Thid{\'e}, Antonio
  Bianchini, and Filippo Romanato.
\newblock Encoding many channels on the same frequency through radio vorticity:
  first experimental test.
\newblock {\em New Journal of Physics}, 14(3):033001, 2012.

\bibitem{tamburini2015tripling}
F~Tamburini, E~Mari, G~Parisi, F~Spinello, M~Oldoni, RA~Ravanelli, P~Coassini,
  Carlo~G Someda, B~Thid{\'e}, and F~Romanato.
\newblock Tripling the capacity of a point-to-point radio link by using
  electromagnetic vortices.
\newblock {\em Radio Science}, 2015.

\bibitem{edfors2012orbital}
Ove Edfors and Anders~J Johansson.
\newblock Is orbital angular momentum (oam) based radio communication an
  unexploited area?
\newblock {\em Antennas and Propagation, IEEE Transactions on},
  60(2):1126--1131, 2012.

\bibitem{xie2015performance}
Guodong Xie, Long Li, Yongxiong Ren, Hao Huang, Yan Yan, Nisar Ahmed, Zhe Zhao,
  Martin~PJ Lavery, Nima Ashrafi, Solyman Ashrafi, et~al.
\newblock Performance metrics and design considerations for a free-space
  optical orbital-angular-momentum--multiplexed communication link.
\newblock {\em Optica}, 2(4):357--365, 2015.

\bibitem{zhang2013restriction}
Yingjie Zhang, Wei Feng, and Ning Ge.
\newblock On the restriction of utilizing orbital angular momentum in radio
  communications.
\newblock In {\em Communications and Networking in China (CHINACOM), 2013 8th
  International ICST Conference on}, pages 271--275. IEEE, 2013.

\bibitem{yan2014high}
Yan Yan, Guodong Xie, Martin~PJ Lavery, Hao Huang, Nisar Ahmed, Changjing Bao,
  Yongxiong Ren, Yinwen Cao, Long Li, Zhe Zhao, et~al.
\newblock High-capacity millimetre-wave communications with orbital angular
  momentum multiplexing.
\newblock {\em Nature communications}, 5, 2014.

\bibitem{mohammadi2010orbital}
Siavoush~Mohaghegh Mohammadi, Lars~KS Daldorff, Jan~ES Bergman, Roger~L
  Karlsson, Bo~Thid{\'e}, Kamyar Forozesh, Tobia~D Carozzi, and Brett Isham.
\newblock Orbital angular momentum in radio—a system study.
\newblock {\em Antennas and Propagation, IEEE Transactions on}, 58(2):565--572,
  2010.

\bibitem{tennant2012generation}
Alan Tennant and Ben Allen.
\newblock Generation of oam radio waves using circular time-switched array
  antenna.
\newblock {\em Electronics letters}, 48(21):1365--1366, 2012.

\bibitem{berry2000making}
Michael Berry.
\newblock Making waves in physics.
\newblock {\em Nature}, 403(6765):21--21, 2000.

\bibitem{Trinder2005}
Julian~Richard Trinder.
\newblock Parabolic reflector.
\newblock {\em patent}, WO 2005/069443, 07 2005.

\bibitem{schemmel2014modular}
Peter Schemmel, Giampaolo Pisano, and Bruno Maffei.
\newblock A modular spiral phase plate design for orbital angular momentum
  generation at millimetre wavelengths.
\newblock {\em Optics express}, 22(12):14712--14726, 2014.

\bibitem{gao2013generating}
Xinlu Gao, Shanguo Huang, Jing Zhou, Yongfeng Wei, Chao Gao, Xukai Zhang, and
  Wanyi Gu.
\newblock Generating, multiplexing/demultiplexing and receiving the orbital
  angular momentum of radio frequency signals using an optical true time delay
  unit.
\newblock {\em Journal of Optics}, 15(10):105401, 2013.

\bibitem{mahmouli2012orbital}
Fariborz~Eslampanahi Mahmouli and Stuart Walker.
\newblock Orbital angular momentum generation in a 60ghz wireless radio
  channel.
\newblock In {\em Telecommunications Forum (TELFOR), 2012 20th}, pages
  315--318. IEEE, 2012.

\bibitem{balanis2005antenna}
Constantine~A Balanis.
\newblock {\em Antenna theory: analysis and design}, volume~1.
\newblock John Wiley \& Sons, 2005.

\bibitem{bai2013generation}
Qiang Bai, Alan Tennant, Ben Allen, and Masood~Ur Rehman.
\newblock Generation of orbital angular momentum (oam) radio beams with phased
  patch array.
\newblock 2013.

\bibitem{wei2015generation}
Wenlong Wei, Kourosh Mahdjoubi, Christian Brousseau, and Olivier Emile.
\newblock Generation of oam waves with circular phase shifter and array of
  patch antennas.
\newblock {\em Electronics Letters}, 51(6):442--443, 2015.

\bibitem{deng2013generation}
Changjiang Deng, Wenhua Chen, Zhijun Zhang, Yue Li, and Zhenghe Feng.
\newblock Generation of oam radio waves using circular vivaldi antenna array.
\newblock {\em International Journal of Antennas and Propagation}, 2013, 2013.

\bibitem{bai2014experimental}
Qiang Bai, Alan Tennant, and Ben Allen.
\newblock Experimental circular phased array for generating oam radio beams.
\newblock {\em Electronics Letters}, 50(20):1414--1415, 2014.

\bibitem{someda2006electromagnetic}
Carlo~G Someda.
\newblock {\em Electromagnetic waves}.
\newblock CRC Press, 2006.

\bibitem{torner2005digital}
Lluis Torner, Juan Torres, and Silvia Carrasco.
\newblock Digital spiral imaging.
\newblock {\em Optics Express}, 13(3):873--881, 2005.

\bibitem{ricci2012instability}
F~Ricci, W~L{\"o}ffler, and MP~van Exter.
\newblock Instability of higher-order optical vortices analyzed with a
  multi-pinhole interferometer.
\newblock {\em Optics express}, 20(20):22961--22975, 2012.

\bibitem{hampton2013introduction}
Jerry~R Hampton.
\newblock {\em Introduction to MIMO Communications}.
\newblock Cambridge University Press, 2013.

\bibitem{parisi2014manipulating}
G~Parisi, E~Mari, F~Spinello, F~Romanato, and F~Tamburini.
\newblock Manipulating intensity and phase distribution of composite
  laguerre-gaussian beams.
\newblock {\em Optics Express}, 22(14):17135--17146, 2014.

\bibitem{bozinovic2013terabit}
Nenad Bozinovic, Yang Yue, Yongxiong Ren, Moshe Tur, Poul Kristensen, Hao
  Huang, Alan~E Willner, and Siddharth Ramachandran.
\newblock Terabit-scale orbital angular momentum mode division multiplexing in
  fibers.
\newblock {\em Science}, 340(6140):1545--1548, 2013.

\end{thebibliography}

\end{document}